\documentclass[11pt]{article}
\usepackage[totalwidth=460truept,totalheight=600truept]{geometry}
\usepackage{latexsym,amssymb,amsmath,graphicx,slashed}
\usepackage{dsfont}
\usepackage[T1]{fontenc}

\def\theequation{\arabic{section}.\arabic{equation}}

\renewcommand{\theequation}{\thesection.\arabic{equation}}
\global\arraycolsep=1truept

\begin{document}

\bigskip

\hfill IFUP-TH 2010/46

\vskip 1.4truecm

\begin{center}
{\huge \textbf{Vacuum Cherenkov Radiation }}

\vskip .4truecm

{\huge \textbf{In Quantum Electrodynamics }}

\vskip .4truecm

{\huge \textbf{With High-Energy Lorentz Violation }}

\vskip 1.5truecm

\textsl{Damiano Anselmi }$^{a,b}$\textsl{\ and {Martina Taiuti} }$^{b,c}$

\vskip .2truecm

$^{a}$\textit{Institute of High Energy Physics, Chinese Academy of Sciences,}

\textit{19 (B) Yuquanlu, Shijingshanqu, Beijing 10049, China,}

\vskip .2truecm

$^{b}$\textit{Dipartimento di Fisica ``Enrico Fermi'', Universit\`{a} di
Pisa, }

\textit{Largo B. Pontecorvo 3, I-56127 Pisa, Italy,}

\vskip .2truecm

$^{c}$\textit{INFN, Sezione di Pisa, }

\textit{Largo B. Pontecorvo 3, I-56127 Pisa, Italy}

\vskip .2truecm

damiano.anselmi@df.unipi.it, martina.taiuti@df.unipi.it\textit{\ }

\textsl{\textit{\vskip 2truecm }}

\textsl{\textit{\textbf{{Abstract} }}}
\end{center}

{\small We study phenomena predicted by a renormalizable, CPT\ invariant
extension of the Standard Model that contains higher-dimensional operators
and violates Lorentz symmetry explicitly at energies greater than some scale 
}$\Lambda _{L}${\small . In particular, we consider the Cherenkov radiation
in vacuo. In a rather general class of dispersion relations, there exists an
energy threshold above which radiation is emitted. The threshold is enhanced
in composite particles by a sort of kinematic screening mechanism. We study
the energy loss and compare the predictions of our model with known
experimental bounds on Lorentz violating parameters and observations of
ultrahigh-energy cosmic rays. We argue that the scale of Lorentz violation }$%
\Lambda _{L}${\small \ (with preserved CPT invariance) can be smaller than
the Planck scale, actually as small as 10}$^{14}$-{\small 10}$^{15}${\small %
GeV. Our model also predicts the Cherenkov radiation of neutral particles.}

\vskip 1truecm

\vfill\eject

\section{Introduction}

\setcounter{equation}{0}

Lorentz symmetry is one of the most precise symmetries in nature \cite
{kostelecky}. Nevertheless, the possibility that it might be violated at
high energies or large distances is still open and has been widely
investigated. If we assume that Lorentz symmetry is not exact, several
phenomena that are otherwise forbidden can occur. Examples are the Cherenkov
radiation in vacuo and the photon decay into an electron-positron pair.
Studying phenomena of this type and comparing predictions with experimental
data, we can look for signs of Lorentz violation and put bounds on the
values of Lorentz violating parameters.

From the theoretical point of view, it is interesting to know that if
Lorentz symmetry is (explicitly) violated at high energies, vertices that
are non-renormalizable by power counting can become renormalizable by a
modified power-counting criterion, which assigns different weights to space
and time \cite{halat}. Consistent models, where the dispersion relations are
modified by higher powers of momenta, can contain operators of higher
dimensions, such as two-scalar--two-fermion vertices and four-fermion
vertices; they are multiplied by inverse powers of some energy $\Lambda _{L}$%
, which can be interpreted as the scale of Lorentz violation. Lorentz
violating gauge theories \cite{LVgauge1,LVgauge2} can be formulated, as well
as Lorentz violating extensions of the Standard Model \cite{lvsm,noh}, which
we call, for brevity, LVSM. In the common perturbative framework, these
theories are unitary, local, polynomial and causal.

Various phenomena that are forbidden in Lorentz invariant theories, but
allowed in Lorentz violating ones, have been studied in the literature,
mainly using the modified dispersion relations of low-energy effective
models. Here we plan to study some of those phenomena in the realm of the
LVSM, where the dispersion relations are crucial for renormalizability,
therefore more constrained and valid, in principle, at arbitrarily high
energies (when gravity is switched off). Our purpose is to derive bounds on
the magnitude of $\Lambda _{L}$. We believe that the scale of Lorentz
violation may be smaller than the Planck scale. If this were true, our
understanding of physics around the Planck scale, in particular quantum
gravity, would have to be reconsidered from scratch.

We assume that CPT is preserved (or that it is violated at energies much
larger than $\Lambda _{L}$). The value of $\Lambda _{L}$ originally
suggested in ref. \cite{lvsm} from considerations about neutrino masses and
bounds on proton decay was $\Lambda _{L}\sim 10^{14}$-$10^{15}$GeV. (In the
appendix we briefly review those arguments and the minimal LVSM.) In this
paper we show that such values are indeed compatible with experimental data
on Lorentz violating phenomena.

Experimental bounds on the parameters that multiply higher-dimensional
operators can be read from the tables of Kostelecky and Russell \cite
{kostelecky}. At present, the best results belong to the photon sector, and
concern the quadratic terms 
\[
F_{k\lambda }\partial _{\alpha _{1}}\cdots \partial _{\alpha _{n}}F_{\mu \nu
}. 
\]
In particular, from astrophysical birefringence and astrophysical dispersion
it is found that the coefficients of the terms of dimensions 6 and 8 are
bounded by 
\[
\lesssim 10^{-29}\text{GeV}^{-2}\text{ and }\lesssim 10^{-25}\text{GeV}^{-4}%
\text{,} 
\]
respectively. Interpreting such coefficients as $\sim 1/\Lambda _{L}^{2}$
and $\sim 1/\Lambda _{L}^{4}$ we see that these experimental data are
consistent with our claim that $\Lambda _{L}$ could be as small as $10^{14}$-%
$10^{15}$GeV.

Under some assumptions, ultrahigh-energy cosmic rays have been claimed to
raise the bound on $\Lambda _{L}$ well above the Planck scale \cite{gagnon}.
However, the nature of ultrahigh-energy cosmic rays has not been firmly
established, yet, so it is not obvious how to use them to put unambiguous
bounds on the scale of Lorentz violation. In this paper we give several
scenarios that are consistent with a value of $\Lambda _{L}$ well below the
Planck scale, assuming that ultrahigh-energy cosmic rays are protons or
heavy nuclei. For our purposes, it will be sufficient to restrict to the
minimal QED subsector of the LVSM, which we call LVQED.

We focus on the Cherenkov radiation in vacuo. For a very general class of
dispersion relations we prove that there exists an energy threshold above
which radiation is emitted and below which it is not emitted. Quite
interestingly, the threshold is enhanced in composite particles by a sort of
kinematic screening mechanism. We study the energy loss as a function of
time and prove that in all cases of our interest it is so rapid that the
radiation is practically governed by pure kinematics. Our models also
predict the Cherenkov radiation of neutral particles.

The paper is organized as follows. In section 2 we present the LVQED model
we are going to study and some basic formulas. In section 3 we study the
Cherenkov radiation in the low-energy expansion. From section 4 onwards we
investigate situations where the standard low-energy expansion does not
apply. Some results can be derived using exact dispersion relations. For
other purposes a different kind of expansion can be used. In section 4 we
study kinematic constraints and derive the energy threshold for Cherenkov
radiation. In section 5 we compare two typical scenarios with experimental
data. In section 6 we study composite particles and show that compositeness
favors larger thresholds. In section 7 we discuss the Cherenkov radiation of
neutrons and neutrinos, while section 8 contains our conclusions.

\section{Preliminaries}

\setcounter{equation}{0}

In this section we write the models we are going to study and derive a
general formula for the energy loss per unit time.

The LVQED model we consider is the minimal QED subsector of the LVSM\
recalled in the appendix, to which we refer for the notation. Its Lagrangian
reads 
\begin{eqnarray}
\mathcal{L} &=&\mathcal{L}_{F}+\bar{\psi}\left( i\gamma ^{0}D_{0}+\frac{%
ib_{0}}{\Lambda _{L}^{2}}{\bar{D}\!\!\!\!\slash}\,^{3}+ib_{1}\bar{D}\!\!\!\!%
\slash -m-\frac{b^{\prime }}{\Lambda _{L}}{\bar{D}\!\!\!\!\slash}%
\,^{2}\right) \psi  \label{mink} \\
&&+\frac{e}{\Lambda _{L}}\bar{\psi}\hspace{0.02in}\left( b^{\prime \prime
}\sigma _{ij}F_{ij}+\frac{b_{0}^{\prime }}{\Lambda _{L}}\gamma _{i}\partial
_{j}F_{ij}\right) \psi +ie\frac{b_{0}^{\prime \prime }}{\Lambda _{L}^{2}}%
F_{ij}\left( \bar{\psi}\gamma _{i}\hspace{0.02in}\frac{\overleftrightarrow{{%
\bar{D}}}_{j}}{2}\psi \right) ,  \nonumber
\end{eqnarray}
where the covariant derivative is $D_{\mu }=\partial _{\mu }+ieA_{\mu }$, $%
\sigma _{\mu \nu }=-i[\gamma _{\mu },\gamma _{\nu }]/2$. Moreover, 
\[
\mathcal{L}_{F}=\frac{1}{2}F_{0i}^{2}-\frac{1}{4}F_{ij}\left( \tau _{2}-\tau
_{1}\frac{\bar{\partial}^{2}}{\Lambda _{L}^{2}}+\tau _{0}\frac{(-\bar{%
\partial}^{2})^{2}}{\Lambda _{L}^{4}}\right) F_{ij} 
\]
is the Lagrangian of free photons. The quantization of this theory has been
studied in ref. \cite{taiuti}. More details can be found there, together
with an analysis of its renormalization. The dispersion relations of
fermions and photons are 
\begin{equation}
E(\bar{p}^{2})=\sqrt{\bar{p}^{2}\left( b_{1}+\frac{b_{0}}{\Lambda _{L}^{2}}%
\bar{p}^{2}\right) ^{2}+\left( \frac{b^{\prime }}{\Lambda _{L}}\bar{p}%
^{2}+m\right) ^{2}},\qquad \omega (\bar{k}^{2})=\sqrt{\tau _{2}\bar{k}%
^{2}+\tau _{1}\frac{(\bar{k}^{2})^{2}}{\Lambda _{L}^{2}}+\tau _{0}\frac{(%
\bar{k}^{2})^{3}}{\Lambda _{L}^{4}}},  \label{disprela}
\end{equation}
respectively.

The low-energy limit of (\ref{mink}) is 
\begin{equation}
\mathcal{L}_{\mathrm{low}}=\frac{1}{2}F_{0i}^{2}-\frac{\tau _{2}}{4}%
F_{ij}^{2}+\bar{\psi}\left( i\gamma ^{0}D_{0}+ib_{1}\bar{D}\!\!\!\!\slash %
-m\right) \psi ,  \label{lelow}
\end{equation}
which formally coincides with the Lagrangian of QED in a medium. The
parameters $\tau _{2}$ and $b_{1}$ are related to the dielectric constant $%
\varepsilon $ and the magnetic permeability $\mu $ by the formulas 
\begin{equation}
\tau _{2}=\frac{\varepsilon }{\mu },\qquad b_{1}=\varepsilon .  \label{repla}
\end{equation}
Moreover, 
\[
n=\sqrt{\varepsilon \mu }=\frac{b_{1}}{\sqrt{\tau _{2}}} 
\]
is the refractive index. Performing the replacements (\ref{repla}) and the
rescalings 
\[
x^{i}\rightarrow \varepsilon x^{i},\qquad A_{i}\rightarrow \frac{A_{i}}{%
\varepsilon },\qquad \psi \rightarrow \frac{\psi }{\varepsilon ^{3/2}}, 
\]
in the action of (\ref{lelow}), we obtain the more common Lagrangian 
\begin{equation}
\mathcal{L}_{\mathrm{medium}}=\frac{\varepsilon }{2}F_{0i}^{2}-\frac{1}{4\mu 
}F_{ij}^{2}+\bar{\psi}\left( iD\!\!\!\!\slash -m\right) \psi .
\label{medium}
\end{equation}
We use for (\ref{medium}) the gauge-fixing term of Lorenz type 
\begin{equation}
\mathcal{L}_{\mathrm{GF}}=-\frac{1}{2\mu }(\varepsilon \mu \partial
_{0}A_{0}-\partial _{i}A_{i})^{2}.  \label{gf}
\end{equation}

We assume that a particle above threshold continuously loses its energy
through the process $e\rightarrow e\gamma $. Then the emitted radiation is
made of a large number of low-frequency photons. The particle remains above
threshold, but its energy asymptotically tends to the threshold value. This
conservative assumption is sufficient for our purposes. Indeed, we are going
to show that in all cases we are interested in the energy loss calculated in
this way is so rapid that we can assume that the particle (practically)
reaches the threshold instantaneously. Other types of emission occur. For
example, the model (\ref{mink}) also contains vertices with two fermions and
two or more photons, which allow elementary processes such as $e\rightarrow
e\gamma \gamma $ and $e\rightarrow e\gamma \gamma \gamma $. It may also
happen \cite{altschul} that the particle loses most of its energy emitting a
single sufficiently energetic photon, or a finite number of photons. Then
the deceleration is not continuous. These effects can increase the rate of
energy loss, but do not affect the conclusions of this paper.

It is convenient to derive a general formula for the energy loss per unit
time without making assumptions on the dispersion relations. It will be
applied to both (\ref{mink}) and (\ref{lelow}). Consider a charged fermion
of energy $E$ and momentum $p$ emitting a photon of frequency $\omega $.
Call $E^{\prime }$ and $p^{\prime }$ the energy and momentum of the fermion
after emission. The expression of the differential width is 
\begin{equation}
\mathrm{d}\Gamma =\frac{1}{2E}\overline{|\mathcal{M}|^{2}}(2\pi )\delta
(E-\omega -E^{\prime })(2\pi )^{3}\delta ^{3}(\mathbf{p}-\mathbf{k}-\mathbf{p%
}^{\prime })\frac{\mathrm{d}^{3}\mathbf{k}}{2\omega (2\pi )^{3}}\frac{%
\mathrm{d}^{3}\mathbf{p}^{\prime }}{2E^{\prime }(2\pi )^{3}},  \label{gam}
\end{equation}
where $\overline{|\mathcal{M}|^{2}}$ is the squared modulus of the
transition amplitude, summed over the final states and averaged over the
initial states.

As usual, the integral over $\mathbf{p}^{\prime }$ is done eliminating the
delta function associated with momentum conservation. The surviving integral
is reduced to an integral over $\omega $ and $u=\cos \theta $, $\theta $
being the angle between the momentum of the incoming fermion and the
momentum of the emitted photon. Next, the delta function of energy
conservation can be used to perform the $u$-integral. It gives $u$ as a
function of $p$ and $k$. Finally, the condition $|u(p,k)|\leqslant 1$
determines the range of the final $k$-integration. We find 
\begin{equation}
\frac{\mathrm{d}\Gamma }{\mathrm{d}\omega }=\frac{\overline{|\mathcal{M}|^{2}%
}}{16\pi Ep}\frac{k}{\omega }\frac{\mathrm{d}k}{\mathrm{d}\omega }%
\sum_{u^{*}}\frac{1}{\left| \frac{E^{\prime }}{p^{\prime }}\frac{\mathrm{d}%
E^{\prime \hspace{0.01in}}}{\mathrm{d}p^{\prime }}\right| _{u=u^{*}}},
\label{diffw}
\end{equation}
where the sum is over the solutions $u^{*}(p,k)$ to the condition of energy
conservation. In the case of (\ref{medium}), the solution is unique.
Instead, the dispersion relations of our Lorentz violating models admit
multiple solutions, in general. Yet, the solution remains unique under quite
reasonable assumptions (see section 4). In this case the $k$-range is of the
standard form $0\leqslant k\leqslant k_{\max }$, for some $k_{\max }$.

The differential width can be used to calculate the energy loss per unit
time, using the formula 
\begin{equation}
\frac{\mathrm{d}E}{\mathrm{d}t}=-\int_{0}^{\omega _{\text{max}}}\omega \frac{%
\mathrm{d}\Gamma }{\mathrm{d}\omega }\mathrm{d}\omega ,  \label{enloss}
\end{equation}
where $\omega _{\max }=\omega (k_{\max }^{2})$.

\section{Cherenkov radiation in QED}

\setcounter{equation}{0}

In this section we study the energy loss of charged particles in empty space
due to the violation of Lorentz symmetry, first in the low-energy theory (%
\ref{lelow}) and later in the low energy expansion of the complete theory (%
\ref{mink}). We then apply our formulas to ultrahigh-energy cosmic rays. The
Cherenkov radiation in vacuo has been studied by various authors \cite
{altschul,glashow,vari,klinkhamer}. Some results of this section are already
available in the literature, others are new.

The Cherenkov radiation occurs if $n>1$, which we assume here. (If $n<1$ a
sufficiently energetic photon can decay into an electron-positron pair, see
for example \cite{glashow}. However, we are not going to study that
phenomenon in this paper). We use the notation (\ref{medium}) and work out
exact formulas without assuming that $n$ is close to 1, so our results can
be also applied to the Cherenkov radiation of charged particles in true
media.

The propagators derived from (\ref{medium}) and (\ref{gf}) are 
\[
\langle A_{\mu }(k)\hspace{0.01in}A_{\nu }(-k)\rangle =\frac{i}{\varepsilon }%
\frac{\text{diag}\left( -1/n^{2},\mathds{1}\right) }{\omega ^{2}-(\mathbf{k}%
^{2}/n^{2})+i0},\qquad \langle \psi (p)\hspace{0.01in}\bar{\psi}(-p)\rangle
=i\frac{\slashed{p} +m}{p^{2}-m^{2}+i0}, 
\]
where $k=(\omega ,\mathbf{k})$. From these expressions we can read the
formulas for the sums over polarization states: 
\begin{equation}
\sum_{\lambda }\varepsilon _{\mu }^{(\lambda )}\varepsilon _{\nu }^{(\lambda
)*}=\frac{1}{\varepsilon }\text{diag}\left( -1/n^{2},\mathds{1}\right)
,\qquad \sum_{s}u_{s}(p)\bar{u}_{s}(p)=\slashed{p} +m,\qquad
\sum_{s}v_{s}(p)\bar{v}_{s}(p)=\slashed{p} -m.  \label{sumfin}
\end{equation}
After a small amount of work we find that formula (\ref{diffw}) gives 
\begin{equation}
\frac{\mathrm{d}\Gamma }{\mathrm{d}\omega }=\frac{\mu \alpha }{2Ep}\left\{ 
\frac{n^{2}-1}{n^{2}}\left[ 2E(E-\omega )+\frac{\omega ^{2}}{2}%
(n^{2}+1)\right] -2m^{2}\right\} ,  \label{guo}
\end{equation}
with 
\[
\omega \leqslant \omega _{\text{max}}=\frac{2(np-E)}{n^{2}-1},\qquad \frac{1%
}{n}\leqslant v\equiv \frac{p}{E}<1. 
\]
In the limit $v\ll 1$, $\omega \ll E$, formula (\ref{guo}) agrees with the
classic one, see e.g. \cite{jackson}. The energy loss (\ref{enloss}) per
unit time is 
\begin{equation}
\frac{\mathrm{d}E}{\mathrm{d}t}=-\frac{\alpha m^{2}\mu (nv-1)^{3}P(v)}{%
3n^{2}(n^{2}-1)^{3}v(1-v^{2})},  \label{loss}
\end{equation}
where 
\[
P(x)=3n(3n^{2}-1)x-(5n^{2}+1). 
\]
The result (\ref{loss}) agrees with the one found by Klinkhamer and Schreck
in ref. \cite{klinkhamer}. We can rewrite it as a differential equation for
the velocity as a function of time: 
\begin{equation}
\frac{\mathrm{d}v}{\mathrm{d}t}=-\frac{\alpha m\mu (nv-1)^{3}\sqrt{1-v^{2}}%
P(v)}{3n^{2}(n^{2}-1)^{3}v^{2}}.  \label{velo}
\end{equation}
The energy decreases to the asymptotic limit 
\begin{equation}
E_{\lim }=\frac{mn}{\sqrt{n^{2}-1}},  \label{lime}
\end{equation}
which corresponds to the asymptotic velocity $v_{\lim }=1/n$. Equation (\ref
{velo}) can be integrated around $v=1$, but not around $v_{\lim }$. This
means that a particle with infinite energy radiates to some final energy $%
E_{f}=m/\sqrt{1-v_{f}^{2}}$ in a finite amount of time $t(n,E_{f})$, but
reaches the energy limit (\ref{lime}) only after an infinite amount of time: 
$t(n,E_{\lim })=\infty $.

\paragraph{Radiation time}

Solving (\ref{velo}) we find 
\begin{eqnarray}
t(n,E_{f}) &=&\frac{3n(n^{2}-1)(3-nv_{f})}{16\alpha E_{f}\mu (nv_{f}-1)^{2}}+%
\frac{3(25n^{4}+14n^{2}-3)}{64\alpha m\mu \sqrt{n^{2}-1}}\ln \frac{n-v_{f}+%
\sqrt{(n^{2}-1)(1-v_{f}^{2})}}{nv_{f}-1}  \nonumber \\
&&-\frac{9(3n^{2}-1)(5n^{2}+1)^{2}}{64\alpha m\mu \sqrt{n^{2}-1}\sqrt{%
P_{+}P_{-}}}\ln \frac{v_{f}P(1/v_{f})+\sqrt{P_{+}P_{-}(n^{2}-1)(1-v_{f}^{2})}%
}{P(v_{f})},  \label{tf}
\end{eqnarray}
where 
\[
P_{\pm }=9n^{2}\pm 4n+1. 
\]

Plotting (\ref{tf}) for various values of $n$ close to 1, we can see that
the energy decrease has a regular shape (see Fig. 1). For all our practical
purposes the particle loses ``all'' its energy during some finite effective
radiation time. However, since the decay is not exponential, the radiation
time must be defined in an unconventional way. 
\begin{figure}[tbp]
\begin{center}
\includegraphics[width=4truein,height=2truein]{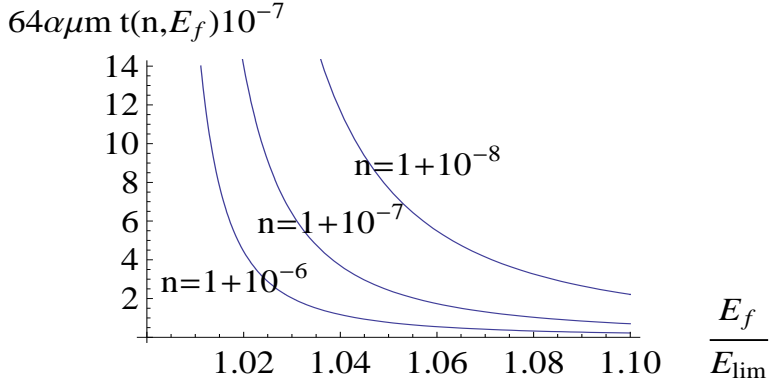}
\end{center}
\caption{Radiation time}
\end{figure}

Assume that the maximum observed energy of a certain class of particles is $%
E_{\text{obs}}\gg m$. Then, if we knew that $E_{\text{obs}}\leqslant E_{\lim
}$ we would obtain the bound 
\begin{equation}
n\leqslant \frac{1}{\sqrt{1-\frac{m^{2}}{E_{\text{obs}}^{2}}}}.
\label{bestb}
\end{equation}
Since we cannot exclude that our $E_{\text{obs}}$ is greater than $E_{\lim }$%
, we must content ourselves with a worse bound. However, we can show that
the decay is so fast that the ``worse'' bound is for all practical purposes
as good as (\ref{bestb}).

We consider ultrahigh-energy cosmic rays, for which we take the highest
observed energy $E_{\text{obs}}=3\cdot 10^{11}$GeV \cite{francesi}. In most
part of this paper, we assume that they are protons or iron atoms moving in
empty space. As far as the fine-structure constant $\alpha $ is concerned,
we use the value $1/116$, namely the Standard-Model value of the running
coupling at $E_{\text{obs}}$, calculated using the beta functions of ref. 
\cite{jones}, the value of $\alpha (M_{z})$ of \cite{mz} and the values of $%
M_{Z}$ and $\sin \theta _{W}(M_{Z})$ from Particle Data Group \cite{pdg}. We
neglect the running of $\alpha $ from $E_{\text{obs}}$ to $\Lambda _{L}$,
because it does not affect out estimates very much. Indeed, $\alpha \sim
1/113$ at $10^{14}$GeV, and $\alpha \sim 1/106$ at $10^{19}$GeV.

Writing $E_{\text{obs}}=rE_{\lim }$, with $r>1$, we have 
\[
n(r)=\frac{1}{\sqrt{1-\frac{m^{2}r^{2}}{E_{\text{obs}}^{2}}}}. 
\]
The age of ultrahigh-energy cosmic rays cannot exceed 
\[
t_{f}(r)=t(n(r),E_{\text{obs}}), 
\]
since when they were created they certainly had a finite energy. Plotting $%
t_{f}(r)$, we see that it is a decreasing function of $r$ and tends to
infinity for $r\rightarrow 1^{+}$. If ultrahigh-energy primaries are
protons, it is easy to check that for $r^{2}=2$ and $\mu =1$, for example,
the time $t_{f}$ is just $\sim $10$^{-10}$ seconds, which means that the
particle loses all its energy down to $E_{\text{obs}}$ in a few centimeters.
Since it certainly covers larger distances, we must have $r<\sqrt{2}$,
therefore 
\[
n\sim 1+\frac{r^{2}}{2}10^{-23}<1+10^{-23}. 
\]
Lowering $r^{2}$ does not improve this bound so much, so we do not need to
struggle to make $r$ as close as we can to $1$ and $t_{f}(r)$ equal to the
age of the Universe (or the time of some intergalactic travel).

If ultrahigh-energy primaries are iron atoms we obtain the weaker bound 
\[
n<1+3\cdot 10^{-20}, 
\]
and $t_{f}$ $\sim 4\cdot $10$^{-14}\sec $.

In summary, for our purposes the energy loss is so rapid that we do not make
any relevant mistake if we use (\ref{bestb}).

\paragraph{$1/\Lambda _{L}$-corrections}

Our model (\ref{mink}) predicts corrections to the results found above,
which can be calculated expanding in powers of $m/\Lambda _{L}$. To
illustrate integrability properties we consider $\mathrm{d}t/\mathrm{d}v$,
instead of $\mathrm{d}v/\mathrm{d}t$. The first correction to $\mathrm{d}t/%
\mathrm{d}v$ is 
\begin{equation}
\Delta \frac{\mathrm{d}t}{\mathrm{d}v}=\frac{3\mu
v^{2}(n^{2}-1)^{2}(48b^{\prime \prime }n^{4}(nv-1)^{2}+b^{\prime }P_{2}(v))}{%
\alpha \Lambda _{L}n^{2}(nv-1)^{4}P(v)^{2}\sqrt{1-v^{2}}},  \label{firstcorr}
\end{equation}
where 
\[
P_{2}(x)=-3n^{2}(3n^{4}+8n^{2}-3)x^{2}+2n(23n^{4}+1)x-25n^{4}+1 
\]
and $v$ still stands for the uncorrected expression 
\begin{equation}
v=\sqrt{1-\frac{m^{2}}{E^{2}}.}  \label{vE}
\end{equation}
We see that $\Delta (\mathrm{d}t/\mathrm{d}v)$ can be integrated
analytically from $v=1$ to any $v_{f}$ greater than $v_{\lim }$. We do not
report the lengthy result here. On the other hand, higher corrections to $%
\mathrm{d}t/\mathrm{d}v$ cannot be integrated around $v=1$, because they
contain factors $(1-v^{2})^{k}$ with $k>1$ in the denominator.

The effects of $1/\Lambda _{L}$- corrections compete with those of $n-1$, so
the expansion in powers of $1/\Lambda _{L}$ is meaningful only if $n$ is not
too close to one. In this section we have assumed that the powers of $n-1$
are dominant. We have seen that the energy loss is so rapid that the
phenomenon is governed by pure kinematics, so corrections such as (\ref
{firstcorr}) are unnecessary. When $n$ is equal to 1, or sufficiently close
to 1, there is no radiation to the zeroth order, or almost none, and we
cannot make a standard low-energy expansion. In the next sections we study
the case when the $1/\Lambda _{L}$-effects are dominant.

\section{Effects of higher space derivatives}

\setcounter{equation}{0}

The LVSM, of which (\ref{mink}) is a subsector, contains terms of higher
dimensions. Under certain conditions those terms are responsible for
Cherenkov radiation in vacuo even if $n$ is exactly one or smaller than one.
Some of them can even cause the radiation of neutral particles. In this
section we begin to study those effects. We first discuss the definition of $%
\Lambda _{L}$ and present our work hypothesis. Then we study the kinematics
of the Cherenkov process.

\paragraph{Definition of $\Lambda _{L}$}

Each term of higher dimension contained in the LVSM can be used to define a
scale of Lorentz violation. Normalizing dimensionless coefficients to one,
we can write a term of this type as 
\[
\frac{1}{\Lambda _{iL}^{d_{i}-4}}\mathcal{O}^{i} 
\]
where $\mathcal{O}^{i}$ is a local operator of dimension $d_{i}>4$
constructed with the fields and their derivatives and $\Lambda _{iL}$ is an
energy scale, which can be regarded as the scale of Lorentz violation
associated with $\mathcal{O}^{i}$.

As far as we know, the values of such $\Lambda _{iL}$'s may significantly
differ from one another. So the question is: which is \textit{the} scale of
Lorentz violation $\Lambda _{L}$? The natural answer is: the smallest $%
\Lambda _{iL}$, namely the smallest energy scale at which the Lorentz
violation may manifest itself. Since the LVSM contains a finite number of
parameters, this definition is meaningful in our approach. Yet, it is a
purely theoretical definition, because no sign of Lorentz violation has been
observed so far.

At the theoretical level, not all parameters of the LVSM are on the same
footing: most of them could be set to zero without affecting the consistency
of the model. Some parameters, on the other hand, must necessarily be
nonzero, because they are crucial for renormalizability. They are the
coefficients that multiply the quadratic terms of the largest dimensions of
each particle: the $\tau _{0}$'s of gauge groups and the $b_{0}$'s of
fermions. In the model (\ref{mink}) the crucial terms are 
\begin{equation}
-\frac{\tau _{0}}{4\Lambda _{L}^{4}}F_{ij}(-\bar{\partial}%
^{2})^{2}F_{ij},\qquad \frac{ib_{0}}{\Lambda _{L}^{2}}\bar{\psi}{\bar{D}%
\!\!\!\!\slash}\,^{3}\psi ,  \label{type}
\end{equation}
while parameters such as $\tau _{1}$, $\tau _{2}-1$, $b_{1}-1$, etc. are not
crucial.

We would like to set the noncrucial parameters to zero, to better isolate
the effects of the crucial ones. However, we have to check whether this is
consistent with renormalization.

We can distinguish parameters according to the dimensions of the operators
they multiply, specifically their \textit{level} $d_{i}-4$. Renormalization
mixing can equalize the orders of magnitude of parameters belonging to the
same level. It can also have important effects on parameters of higher
levels, but not so much on those of lower levels. Indeed, the beta functions
of parameters belonging to lower levels receive contributions that are
suppressed by powers of $m/\Lambda _{L}$. For definiteness, consider the
subset of couplings $\tau _{0,1}$, $b_{0}$. Observe that $\tau _{0}$ is the
only parameter of level 4, while $b_{0}$ and $\tau _{1}$ are of level 2. The
beta functions have structures 
\[
\beta _{\tau _{0}}\sim \alpha \tau _{0}+b_{0}^{2}+b_{0}\tau _{1}+\tau
_{1}^{2},\qquad \beta _{b_{0}}\sim \alpha \tau _{1}+\alpha b_{0}+\frac{m^{2}%
}{\Lambda _{L}^{2}}\tau _{0},\qquad \beta _{\tau _{1}}\sim \alpha \tau
_{1}+\alpha b_{0}+\frac{m^{2}}{\Lambda _{L}^{2}}\tau _{0}, 
\]
where we have written only the first contributing terms of each type. Since $%
m/\Lambda _{L}$ is around $10^{-12}$, at worst, parameters of lower levels
can be consistently set to have much smaller values than parameters of
higher levels. In our case, $\tau _{0}$ will be of order 1 and $b_{0}$ will
be of order one or much smaller than one.

This analysis is sufficient to justify the first scenario studied in the
next section. Sometimes, however, it is interesting to study cases where
particular relations among parameters of the same level hold, but then the
effects of renormalization on those relations need to be studied carefully.
The second scenario studied in the next section provides an example of this.

To summarize, the parameters of the Lorentz violating extended Standard
Model can be arranged according to a hierarchy of conceptual importance,
which may or may not correspond to a hierarchy of magnitude. We take it as a
work hypothesis to organize our analysis. We assume that the absolute values
of the non-crucial parameters are as small as possible, and concentrate on
the crucial ones.

The values of the crucial parameters themselves can significantly differ
from one another. The largest of them defines $\Lambda _{L}$. For example,
if the scale of Lorentz violation $\Lambda _{L}$ is defined by the crucial
term belonging to the photon sector, namely 
\begin{equation}
-\frac{1}{4}F_{ij}\frac{(-\bar{\partial}^{2})^{2}}{\Lambda _{L}^{4}}F_{ij},
\label{normali}
\end{equation}
then we can set $\tau _{0}=1$ for the photon, and assume that all other $%
\tau _{0}$'s, and the $b_{0}$'s, are not greater than 1. This choice sounds
reasonable, indeed, because the photon sector contains the best measured
parameters among those multiplying operators of higher dimensions \cite
{kostelecky}. Under these assumptions, our plan is to study how small the
parameters $b_{0}$'s have to be to explain data, in particular
ultrahigh-energy cosmic rays.

In the rest of this section we study the kinematics of a large class of
dispersion relations. In particular, we study the threshold for Cherenkov
radiation and the range of frequencies of the emitted photon.

\paragraph{General kinematics}

As before, $p$ denotes the momentum of the incoming fermion, $k$ is the
momentum of the emitted photon, $u=\cos \theta $ and $\theta $ is the angle
between the trajectory of the incoming fermion and the photon.

We just assume that at $p,k\neq 0$ the dispersion relations $E(p)$ and $%
\omega (k)$ are non-negative, have positive first derivatives (namely
velocities are always positive) and non-negative second derivatives, and
that at least one dispersion relation is convex: 
\begin{equation}
E\geqslant 0,\qquad \omega \geqslant 0,\qquad \frac{\mathrm{d}E}{\mathrm{d}p}%
>0,\qquad \frac{\mathrm{d}\omega }{\mathrm{d}k}>0,\qquad \frac{\mathrm{d}%
^{2}E}{\mathrm{d}p^{2}}>0,\qquad \frac{\mathrm{d}^{2}\omega }{\mathrm{d}k^{2}%
}\geqslant 0.  \label{ipotesi}
\end{equation}
These properties are obeyed by the usual relativistic and non-relativistic
dispersion relations. In relativistic dispersion relations convexity holds
any time the mass is non-vanishing.

Energy and momentum conservations imply 
\begin{equation}
E(p)=\omega (k)+E(p^{\prime }),\qquad p^{\prime }=\sqrt{p^{2}+k^{2}-2pku}.
\label{ros}
\end{equation}
The condition (\ref{ros}) is involved, but some inequalities that are useful
for the calculation can be derived straightforwardly. For example, we have 
\begin{equation}
k<2p.  \label{rough}
\end{equation}
This information is quite redundant (the precise $k$-range is determined
below), but enough for the moment. It can be proved observing that $%
E(p)-E(p^{\prime })\geqslant 0$ implies $p\geqslant p^{\prime }$, by the
monotonicity of $E(p)$, while $k\geqslant 2p$ would give $p^{\prime
}\geqslant p$ (using $u\leqslant 1$).

Next, consider the condition of energy conservation (\ref{ros}) in the $%
(k,p^{\prime })$-plane and call its solution $p^{\prime }(k)$. For given $k$
the equation for $p^{\prime }$ reads $E(p^{\prime })$=constant. Since the
function $E(p^{\prime })$ is monotonic, the solution $p^{\prime }(k)$, when
it exists, is unique. Second, $p^{\prime }=|\mathbf{p}-\mathbf{k}|$ and $%
p^{\prime }\leqslant p$ tell us that we must focus on the region 
\[
|p-k|\leqslant p^{\prime }\leqslant p. 
\]
Third, $k=0$, $p^{\prime }=p$ is a solution of (\ref{ros}), so $p^{\prime
}(0)=p$.

\begin{figure}[tbp]
\begin{center}
\includegraphics[width=3.2truein,height=1.6truein]{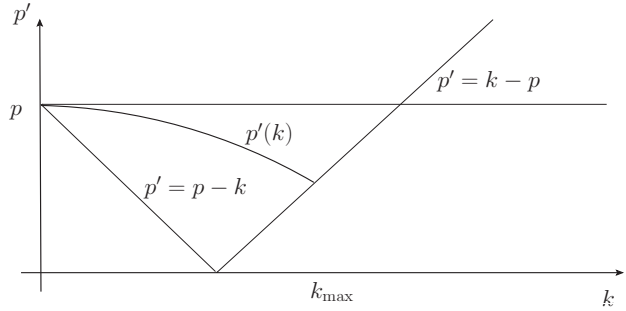}
\end{center}
\caption{General kinematics}
\end{figure}

Finally, $p^{\prime }(k)$ is monotonically decreasing and concave. These
properties are proved differentiating (\ref{ros}) with respect to $k$ once
and twice and using (\ref{ipotesi}): we find, at $k\neq 0$, 
\[
\frac{\mathrm{d}p^{\prime }}{\mathrm{d}k}=-\frac{\mathrm{d}\omega }{\mathrm{d%
}k}\left( \left. \frac{\mathrm{d}E}{\mathrm{d}p}\right| _{p^{\prime
}}\right) ^{-1}<0,\qquad \frac{\mathrm{d}^{2}p^{\prime }}{\mathrm{d}k^{2}}%
=-\left[ \frac{\mathrm{d}^{2}\omega }{\mathrm{d}k^{2}}+\left. \frac{\mathrm{d%
}^{2}E}{\mathrm{d}p^{2}}\right| _{p^{\prime }}\left( \frac{\mathrm{d}%
p^{\prime }}{\mathrm{d}k}\right) ^{2}\right] \left( \left. \frac{\mathrm{d}E%
}{\mathrm{d}p}\right| _{p^{\prime }}\right) ^{-1}<0. 
\]
Here the notation $\left. X\right|_{p^{\prime}}$ specifies where the function $X$ has to be 
evaluated, namely $\left. X\right|_{p^{\prime}}=X(p^{\prime})$.

Using these pieces of information, we can draw the picture of Fig. 2. We see
that a non-trivial range of solutions exists if and only if the first
derivative of $p^{\prime }(k)$ is smaller than one in modulus at $k=0$,
namely 
\begin{equation}
-\left. \frac{\mathrm{d}p^{\prime }}{\mathrm{d}k}\right| _{0}<1\qquad 
\mathrm{or, \, equivalently,} \text{\qquad }\left. \frac{\mathrm{d}\omega }{%
\mathrm{d}k}\right| _{0}<\frac{\mathrm{d}E}{\mathrm{d}p},  \label{fina}
\end{equation}
which means that the velocity of the charged particle must be greater than a
certain threshold determined by the photon dispersion relation, as in the
usual case. Moreover, the $k$-range is the segment 
\begin{equation}
0\leqslant k\leqslant k_{\max }(p),  \label{krange}
\end{equation}
where $k_{\max }(p)$ is the solution of $p^{\prime }(k_{\max })=|p-k_{\max
}| $, namely it is obtained from the forward emission $u=1$.

Observe that the condition (\ref{fina}) does not depend on most parameters
of $\omega (k)$. When the dispersion relations are (\ref{disprela}), (\ref
{fina}) does not depend on $\tau _{0}$ and $\tau _{1}$, but only $\tau _{2}$
and the parameters of the fermion dispersion relation.

\section{Typical scenarios}

\setcounter{equation}{0}

In this section we study two scenarios and their compatibility with the
observation of ultrahigh-energy cosmic rays and other experimental data. Our
purpose is to show that there exist reasonable scenarios where the scale of
Lorentz violation is smaller than the Planck scale.

From the propagators of (\ref{mink}), given in \cite{taiuti}, we can derive
the following sums over polarization states, to be used in formula (\ref
{diffw}): 
\begin{eqnarray}
\sum_{\lambda }\varepsilon _{\mu }^{(\lambda )}\varepsilon _{\nu }^{(\lambda
)*} &=&\text{diag}\left( -\omega ^{2}(\bar{k}^{2})/\bar{k}^{2},\mathds{1}%
\right) ,\qquad \sum_{s}u_{s}(p)\bar{u}_{s}(p)=\slashed{p} +m+\slashed{\bar{p}}
\left( b_{1}-1+\frac{b_{0}}{\Lambda _{L}^{2}}\bar{p}^{2}\right) +%
\frac{b^{\prime }}{\Lambda _{L}}\bar{p}^{2},  \nonumber \\
&&\sum_{s}v_{s}(p)\bar{v}_{s}(p)=\slashed{p} -m+\slashed{\bar{p}} \left(
b_{1}-1+\frac{b_{0}}{\Lambda _{L}^{2}}\bar{p}^{2}\right) -\frac{b^{\prime }}{%
\Lambda _{L}}\bar{p}^{2}.  \label{ave}
\end{eqnarray}

\subsection{First scenario}

In the first example we set all non-crucial parameters to zero apart from
the mass, namely we assume 
\begin{equation}
\tau _{2}=1,\qquad \tau _{1}=0,\qquad b_{1}=1,\qquad b^{\prime }=0,\qquad
b_{0}>0.  \label{seconda}
\end{equation}
The results do not depend very much on the value of $\tau _{1}$ (see comment
on this at the end of this section). We have the dispersion relations 
\begin{equation}
E(p^{2})=\sqrt{m^{2}+p^{2}\left( 1+\frac{b_{0}p^{2}}{\Lambda _{L}^{2}}%
\right) ^{2}},\qquad \omega (k^{2})=\sqrt{k^{2}+\tau _{0}\frac{(k^{2})^{3}}{%
\Lambda _{L}^{4}}}.  \label{dispe2}
\end{equation}
The inequality $b_{0}>0$ is assumed to ensure monotonicity. The condition (%
\ref{fina}) gives 
\begin{equation}
\xi ^{2}\equiv \frac{m^{2}\Lambda _{L}^{2}}{6b_{0}p^{4}}<\left( 1+\frac{%
b_{0}p^{2}}{\Lambda _{L}^{2}}\right) ^{2}\left( 1+\frac{3b_{0}p^{2}}{%
2\Lambda _{L}^{2}}\right) .  \label{fina3}
\end{equation}

We are interested in the case 
\begin{equation}
m\ll p\ll \Lambda _{L},  \label{approx}
\end{equation}
which can help us solve the kinematic constraints in an approximate way.
Specifically, we have 
\[
p\leqslant 3\cdot 10^{11}\text{GeV},\qquad \Lambda _{L}\geqslant 10^{14}%
\text{GeV.} 
\]
Within our approximation the right-hand side of (\ref{fina3}) is practically
1, so the condition for the emission of Cherenkov radiation is 
\begin{equation}
\xi <1,  \label{xi1}
\end{equation}
which can also be expressed as an energy threshold, namely 
\[
E>E_{\lim }\sim \frac{m^{1/2}\Lambda _{L}^{1/2}}{6^{1/4}b_{0}^{1/4}}. 
\]
A particle above threshold radiates and loses energy till it reaches the
limit value $E_{\lim }$.

When $\Lambda _{L}\rightarrow \infty $ at $m$ and $p$ fixed, the condition (%
\ref{ros}) admits no solution, because it reduces to the kinematic relation
of the Lorentz invariant theory. It is more convenient to study the limit $%
\Lambda _{L}\rightarrow \infty $ at $\xi $ and $p$ fixed, because in such a
limit (\ref{ros}) becomes 
\[
p=k+\sqrt{p^{2}+k^{2}-2pku}, 
\]
so its solution is $u=1$, $k\leqslant p$. For $\Lambda _{L}<\infty $ we can
find an approximate solution of the form 
\begin{equation}
u=1-\varepsilon ,\qquad 0<\varepsilon \ll 1\text{,}  \label{epso}
\end{equation}
with $k$ belonging to a certain range of values that has to be worked out.
We can expand in $\varepsilon $ and $p/\Lambda _{L}$ around the $\Lambda
_{L}=\infty $-solution. We find 
\begin{equation}
\varepsilon =\frac{b_{0}}{pk\Lambda _{L}^{2}}\left\{ (p-k)\left[
p^{3}-(p-k)^{3}\right] -3\xi ^{2}p^{3}k\right\} .  \label{epsilon2}
\end{equation}
We see that $\varepsilon $ is indeed much smaller than one, as needed for
consistency. The $k$-range can be read from the condition $\varepsilon
\geqslant 0$. Plotting the functions appearing in (\ref{epsilon2}) it is
easy to show that for $\xi <1$ a range of the form (\ref{krange}) exists and
has $k_{\max }<p$.

The energy losses (\ref{enloss}) can be worked out starting from the
differential width (\ref{diffw}). For the analysis of ultrahigh-energy
cosmic rays it is sufficient to consider the situations $\xi ^{2}\ll 1$ and $%
1-\xi ^{2}\ll 1$. For $\xi ^{2}\ll 1$ we obtain the range 
\[
0\leqslant k\leqslant p(1-3\xi ^{2}) 
\]
and the energy loss 
\begin{equation}
\left. \frac{\mathrm{d}E}{\mathrm{d}t}\right| _{\xi ^{2}\ll 1}=-\frac{%
11\alpha p^{4}b_{0}}{12\Lambda _{L}^{2}}.  \label{zetapiccolo2}
\end{equation}
For $1-\xi ^{2}\ll 1$ we obtain the range 
\[
0\leqslant k\leqslant \frac{p}{2}(1-\xi ^{2}), 
\]
and the energy loss 
\begin{equation}
\left. \frac{\mathrm{d}E}{\mathrm{d}t}\right| _{1-\xi ^{2}\ll 1}=-\frac{%
\alpha p^{4}\left( 1-\xi ^{2}\right) ^{3}}{4\Lambda _{L}^{2}}b_{0}.
\label{zeta12}
\end{equation}
The exact formulas depend also on $b^{\prime \prime }$. However, we have set 
$b^{\prime \prime }=0$, since $b^{\prime \prime }$ is not in the list of
crucial parameters.

Recall that, since we have used the approximation (\ref{approx}), we cannot
use (\ref{zetapiccolo2}) and (\ref{zeta12}) above $E=\Lambda _{L}$. As in
the case of QED in a medium, the radiating particle takes an infinite amount
of time to reach the energy limit. For our purposes it is sufficient to
calculate the time the particle takes to radiate from energy $\Lambda _{L}$
to, say, $1.3$-$1.1E_{\lim }$. It is not meaningful to approach the energy
limit further, since the energies we are considering are not measured so
precisely.

Now we apply our results to ultrahigh-energy cosmic rays. If $\Lambda
_{L}^{{}}=10^{14}$GeV, protons of $3\cdot 10^{11}$GeV emit Cherenkov
radiation if $b_{0}>1.8\cdot 10^{-19}$. If $\Lambda _{L}^{{}}=10^{14}$GeV
and $b_{0}=1.8\cdot 10^{-19}$ we can use (\ref{zetapiccolo2}) as long as $%
\xi $ is small, for example down to $2E_{\lim }$ ($\xi ^{2}=1/16$). The time
spent to radiate from $\Lambda _{L}$ to $2E_{\lim }$ is 
\[
t_{f}^{\prime }\sim 7\cdot 10^{-12}\text{sec.} 
\]
When the energy approaches $E_{\lim }$ we have to use (\ref{zeta12}). The
particle radiates from energy $2E_{\lim }$ to $1.1E_{\lim }$ in about 
\[
t_{f}^{\prime \prime }\sim 8\cdot 10^{-10}\text{sec.} 
\]
The radiation time $t_{f}=t_{f}^{\prime }+t_{f}^{\prime \prime }$ is too
short to be compatible with the observation of ultrahigh-energy cosmic rays.
Therefore, as in section 2, we may assume that the energy loss down to $%
E_{\lim }$ occurs instantaneously any time it is allowed by kinematics.

Larger values of $b_{0}$ give smaller $t_{f}$'s. For example, if $b_{0}\sim
1 $ and $\Lambda _{L}^{{}}=10^{14}$GeV particles radiate down to $E_{\lim
}=6\cdot 10^{6}$GeV in a much shorter time. Integrating (\ref{zetapiccolo2})
from $\Lambda _{L}$ to $E_{\text{obs}}=3\cdot 10^{11}$GeV, we obtain 
\[
t_{f}^{\prime }\sim 10^{-29}\text{sec}, 
\]
while continuing down to $1.1E_{\lim }$ we have to use both (\ref
{zetapiccolo2}) and (\ref{zeta12}), and get 
\[
t_{f}^{\prime \prime }\sim 2\cdot 10^{-14}\text{sec.} 
\]
Thus, only the values $b_{0}\leqslant 1.8\cdot 10^{-19}$ are consistent with
data if $\Lambda _{L}^{{}}=10^{14}$GeV and the cosmic rays are protons.

The limiting value on $b_{0}$ can be raised increasing $\Lambda _{L}$. For
various values of $\Lambda _{L}^{{}}$the bounds on $b_{0}$ are 
\begin{equation}
b_{0\text{lim}}=1.8\cdot 10^{-19+2k}\qquad \text{for }\Lambda _{L}=10^{14+k}%
\text{GeV}  \label{tbl1}
\end{equation}
and $b_{0\text{lim}}=1$ for $\Lambda _{L}=2.4\cdot 10^{23}$GeV. When $%
\Lambda _{L}$ is varied between $10^{14}$GeV and the Planck scale $t_{f}$
does not change very much.

If the ultrahigh-energy cosmic rays are iron atoms we get the bounds 
\begin{equation}
b_{0\text{lim}}=5.6\cdot 10^{-16+2k}\qquad \text{for }\Lambda _{L}=10^{14+k}%
\text{GeV}  \label{tbl2}
\end{equation}
and $b_{0\text{lim}}=1$ for $\Lambda _{L}=4.2\cdot 10^{21}$GeV.

We have also considered a variant of (\ref{seconda}), with $\tau _{1}=2\sqrt{%
\tau _{0}}$ instead of $\tau _{1}=0$. The radiation times are still too
short and the threshold condition is exactly the same, therefore the bounds (%
\ref{tbl1}) and (\ref{tbl2}) are unchanged.

Most bounds we have found are very small. However, the situation improves if
we take compositeness into account. Before that, we study a second
scenario.

\subsection{Second scenario}

The procedure just used is quite general, and can be used to examine other
cases. We illustrate a second scenario taking the dispersion relations 
\begin{equation}
E(p^{2})=\sqrt{m^{2}+p^{2}+b_{0}^{2}\frac{(p^{2})^{3}}{\Lambda _{L}^{4}}}%
,\qquad \omega (k^{2})=\sqrt{k^{2}+\tau _{0}\frac{(k^{2})^{3}}{\Lambda
_{L}^{4}}},  \label{dispe}
\end{equation}
for the fermion energy and photon frequency. Here we assume that the
parameters of the Lagrangian (\ref{mink}) satisfy 
\begin{equation}
b_{0}=-\frac{b^{\prime \hspace{0.01in}2}}{2b_{1}},\qquad b_{1}=\sqrt{1-2%
\frac{mb^{\prime }}{\Lambda _{L}}},\qquad \tau _{2}=1,\qquad \tau _{1}=0,
\label{aut}
\end{equation}
namely they are such that only the highest powers of momentum, which are the
crucial ones for renormalization, correct the relativistic dispersion
relations.

Here it is convenient to define 
\begin{equation}
\xi \equiv \frac{\Lambda _{L}^{2}m}{\sqrt{5}|b_{0}|p^{3}}.  \label{fina2}
\end{equation}
The approximate condition for emission is again $\xi <1$. The limit energy
is 
\[
E_{\lim }=\frac{\Lambda _{L}^{2/3}m^{1/3}}{5^{1/6}|b_{0}|^{1/3}}. 
\]

Setting $u=1-\varepsilon $ as before and $\zeta \equiv b_{0}^{2}/\tau _{0}$,
we find 
\begin{equation}
\varepsilon \sim \frac{b_{0}^{2}}{2pk\Lambda _{L}^{4}}\left\{ (p-k)\left[
p^{5}-(p-k)^{5}-\frac{k^{5}}{\zeta }\right] -5\xi ^{2}p^{5}k\right\} .
\label{angle}
\end{equation}

For $\zeta \ll 1$ it is sufficient to consider the case $\zeta \ll (1-\xi
^{2})^{3}$, which gives the $k$-range 
\begin{equation}
0\leqslant k\leqslant p(5\zeta (1-\xi ^{2}))^{1/4}.  \label{krange1}
\end{equation}
We find, to the lowest order in $1/\Lambda _{L}$ and $\zeta $ (at fixed $%
\tau _{0}$), the energy loss 
\begin{equation}
\left. \frac{\mathrm{d}E}{\mathrm{d}t}\right| _{\zeta \ll 1,\zeta \ll (1-\xi
^{2})^{3}}=-\frac{5\alpha (1-\xi ^{2})\zeta ^{3/2}\sqrt{\tau _{0}}p^{4}}{%
4\Lambda _{L}^{2}}.  \label{dEsudt}
\end{equation}
As before, we have set $b^{\prime \prime }=0$. Instead $b^{\prime }$ must be
kept, because formulas (\ref{aut}) relate it to $b_{0}$. Note that since $%
b_{1}\sim 1$, $b_{0}$ must be negative. Thus we have $b^{\prime 2}\sim
-2b_{0}=2\sqrt{\zeta \tau _{0}}$.

Formula (\ref{dEsudt}) can be integrated exactly. We obtain 
\begin{equation}
\left. E^{3}(t)\right| _{\zeta \ll 1,\zeta \ll (1-\xi ^{2})^{3}}=E_{\lim
}^{3}\frac{\Lambda _{L}^{3}\cosh (\kappa t)+E_{\lim }^{3}\sinh (\kappa t)}{%
\Lambda _{L}^{3}\sinh (\kappa t)+E_{\lim }^{3}\cosh (\kappa t)},
\label{thesolu}
\end{equation}
where 
\[
\kappa =\frac{3\sqrt{5}}{4}\alpha m\zeta . 
\]
and the initial condition is fixed setting $E(0)=\Lambda _{L}$. Formula (\ref
{thesolu}) allows us to define a radiation time in a familiar way, since it
contains only exponentials. We have 
\begin{equation}
\left. t_{f}\right| _{\zeta \ll 1,\zeta \ll (1-\xi ^{2})^{3}}\sim \frac{1}{%
\kappa },  \label{radti0}
\end{equation}
which is strictly speaking the time taken to reach the energy $\sim
1.1E_{\lim }$ (assuming $\Lambda _{L}\gg E_{\lim }$, which is true in the
cases studied here).

If $\zeta =1$ we need to distinguish the cases $\xi ^{2}\ll 1$ and $1-\xi
^{2}\ll 1$. We find the $k$-ranges 
\[
0\leqslant k\leqslant p(1-\xi ),\qquad 0\leqslant k\leqslant \frac{p}{3}%
(1-\xi ^{2}), 
\]
and the energy losses 
\begin{equation}
\left. \frac{\mathrm{d}E}{\mathrm{d}t}\right| _{\zeta =1,\xi ^{2}\ll 1}=-%
\frac{\alpha p^{4}|b_{0}|}{20\Lambda _{L}^{2}},\qquad \left. \frac{\mathrm{d}%
E}{\mathrm{d}t}\right| _{\zeta =1,\xi ^{2}\sim 1}=-\frac{\alpha (1-\xi
^{2})^{4}|b_{0}|p^{4}}{324\Lambda _{L}^{2}},  \label{dEsudt2}
\end{equation}
respectively.

We take $\tau _{0}=1$, which means that we assume that the scale of Lorentz
violation $\Lambda _{L}$ is defined by the photon sector, precisely by the
first term of (\ref{type}). With $\Lambda _{L}=10^{14}$GeV, protons of $%
3\cdot 10^{11}$GeV emit Cherenkov radiation if $|b_{0}|>1.6\cdot 10^{-7}$.
If we take $b_{0}=-1.6\cdot 10^{-7}$ the approximation $\zeta \ll (1-\xi
^{2})^{3}$ holds in the entire energy range from $\Lambda _{L}$ down to $%
1.1E_{\lim }$. We find that the typical radiation time of the particles
above threshold is 
\[
t_{f}\sim 2\cdot 10^{-9}\text{sec.} 
\]

Again, larger values of $b_{0}$ give smaller $t_{f}$'s. For example, if $%
b_{0}\sim 1$ and $\Lambda _{L}^{{}}=10^{14}$GeV a proton of energy $%
E_{f}=3\cdot 10^{11}$GeV has $\xi \sim 10^{-7}$, so the time it spends to
radiate from energy $\Lambda _{L}$ to the final energy $E_{f}\gg E_{\lim }$
can be calculated using the first formula of (\ref{dEsudt2}). We find 
\begin{equation}
\left. t_{f}\right| _{\zeta =1,\xi ^{2}\ll 1}\sim \frac{40\Lambda _{L}^{2}}{%
3\alpha b^{\prime \hspace{0.01in}2}E_{f}^{3}}.  \label{radti}
\end{equation}
Numerically, taking $b_{0}=-1$ and $b^{\prime }\sim \sqrt{2}$, we have 
\begin{equation}
t_{f}\sim 2\cdot 10^{-28}\text{sec.}  \label{time}
\end{equation}
After this time, the cosmic rays keep radiating till they reach the limit
energy, which is $E_{\lim }\sim 1.6\cdot 10^{9}$GeV. We can use the first
formula of (\ref{dEsudt2}) as long as $\xi ^{2}$ is small, for example down
to $2E_{\lim }$ ($\xi ^{2}=1/64$). The time spent to radiate from $\Lambda
_{L}$ to $2E_{\lim }$ is 
\[
t_{f}^{\prime }\sim 1.6\cdot 10^{-22}\text{sec.} 
\]
When the energy approaches $E_{\lim }$ we have to use the second formula of (%
\ref{dEsudt2}). The particle radiates from energy $2E_{\lim }$ to $%
1.1E_{\lim }$ during 
\begin{equation}
t_{f}^{\prime \prime }\sim 6\cdot 10^{-20}\text{sec,}  \label{timep}
\end{equation}
which is still very short.

Summarizing, we may assume that the energy loss is instantaneous, so only
the values $|b_{0}|\leqslant 1.6\cdot 10^{-7}$ are consistent with data at $%
\Lambda _{L}=10^{14}$GeV. The limiting value on $|b_{0}|$ can be raised
increasing $\Lambda _{L}$ and becomes 1 for $\Lambda _{L}=2.5\cdot 10^{17}$%
GeV. If $b_{0}\sim 1$ and $\Lambda _{L}\geq $ $2.5\cdot 10^{17}$GeV, protons
of $3\cdot 10^{11}$GeV do not emit Cherenkov radiation and can reach the
earth. Protons above threshold have a radiation time $t_{f}^{\prime
}+t_{f}^{\prime \prime }$ of about $6\cdot 10^{-20}$sec. We obtain the
bounds 
\begin{equation}
|b_{0\text{lim}}|=1.6\cdot 10^{-7+2k},\qquad \text{for }\Lambda
_{L}=10^{14+k}\text{GeV,}  \label{table}
\end{equation}
and $|b_{0}|=1$ for $\Lambda _{L}=2.5\cdot 10^{17}$GeV. If the
ultrahigh-energy cosmic rays are instead iron atoms their observation can be
explained, for example, with $|b_{0}|\sim 1$, $|b^{\prime }|\sim \sqrt{2}$
and $\Lambda _{L}\sim $ $3.4\cdot 10^{16}$GeV, or with $|b_{0}|\sim 9\cdot
10^{-6}$, $|b^{\prime }|\sim 4\cdot 10^{-3}$ and $\Lambda _{L}\sim $ $%
10^{14} $GeV.

Since we have assumed that the relations (\ref{aut}) hold, we must check the
compatibility of $b^{\prime }$ and $b_{1}$ with present data. Using $%
b_{1}\sim 1$, $p\ll \Lambda _{L}$ and (\ref{aut}), (\ref{fina2}) we get 
\begin{equation}
-\frac{\Lambda _{L}^{2}m}{\sqrt{5}p^{3}}\leqslant b_{0}<0,\qquad |b^{\prime
}|\leqslant \left( \frac{4}{5}\right) ^{\frac{1}{4}}\frac{\Lambda _{L}m^{1/2}%
}{p^{3/2}},\qquad 1-\left( \frac{4}{5}\right) ^{\frac{1}{4}}\left( \frac{m}{p%
}\right) ^{\frac{3}{2}}\leqslant n\leqslant 1+\left( \frac{4}{5}\right) ^{%
\frac{1}{4}}\left( \frac{m}{p}\right) ^{\frac{3}{2}}.  \label{ines}
\end{equation}
Here $p$ is the largest momentum at which the particle is known not to
radiate and $n=b_{1}/\sqrt{\tau _{2}}$ is the refractive index of the vacuum
``as seen by the proton''. Observe that the bound on the refractive index is
independent of $\Lambda _{L}$, so it cannot be improved changing the scale
of Lorentz violation.

The three inequalities (\ref{ines}) are equivalent to one another. We search
for the largest $|b_{0}|$ compatible with data. If $|b_{0}|$ is not small
enough, $n$ may be too far from one, which may contradict existing bounds.

If $b^{\prime }>0$ we find $1-5\cdot 10^{-18}\leqslant n<1$. At present no
bounds contradict this range \cite{kostelecky}. Instead, if $b^{\prime }<0$
we find $1<n\leqslant 1+5\cdot 10^{-18}$. In this case, a bound exists in
the literature, $n<1+6\cdot 10^{-20}$ \cite{klinkhamer,kostelecky}, but it
cannot be applied here, since it is derived from ultrahigh-energy cosmic
rays themselves, which we are explaining with a different approach. Thus,
the largest $|b_{0}|$ we can take is given by 
\begin{equation}
b_{0}=-\frac{\Lambda _{L}^{2}m}{\sqrt{5}p^{3}}.  \label{bili}
\end{equation}

Now we discuss the consistency of the dispersion relations (\ref{dispe})
with renormalization. The first condition of (\ref{aut}) demands that the
combination

\[
\epsilon \equiv 2b_{1}b_{0}+b^{\prime \hspace{0.01in}2} 
\]
vanish, to ensure that the dispersion relations do not contain terms
proportional to the forth power of momentum. A typical case with $\epsilon $
different from zero is the first scenario already studied. The $b_{0}$%
-bounds of (\ref{tbl1}) and (\ref{tbl2}) tell us how small $\epsilon $ must
be to have compatibility with data, in the cases of protons and iron atoms,
respectively. For example, for protons $\epsilon \sim 4\cdot 10^{-19}$ at $%
\Lambda _{L}=10^{14}$GeV. Instead, the results found in the second scenario
tell us $2b_{1}b_{0}\sim b^{\prime \hspace{0.01in}2}\sim 3.2\cdot 10^{-7}$,
which is 12 orders of magnitude larger!

For the reasons explained in the previous section, we may assume that the
relations (\ref{aut}) are valid at the scale $\Lambda _{L}$, or anyway just
at one energy scale. However, the scale we need to work with is $E_{\text{obs%
}}=3\cdot 10^{11}$GeV. The $b_{0}$- and $b^{\prime }$-runnings contain,
among the others, terms proportional to 
\[
\alpha b_{0}\ln \frac{\Lambda _{L}}{E_{\text{obs}}},\qquad \alpha b^{\prime
}\ln \frac{\Lambda _{L}}{E_{\text{obs}}}. 
\]
So, assuming that the cancellation $\epsilon =0$ occurs at $\Lambda _{L}$ it
will not necessarily occur at $E_{\text{obs}}$, where instead we find 
\[
\epsilon \sim \alpha b_{0}\ln \frac{\Lambda _{L}}{E_{\text{obs}}}. 
\]
The values of (\ref{table}) give a too large $\epsilon $. Therefore,
renormalization forces us to take $b_{0}$-values much smaller than the ones
given in (\ref{table}). Precisely, they are just a factor 
\begin{equation}
\frac{1}{\alpha \ln \frac{\Lambda _{L}}{E_{\text{obs}}}}  \label{orders}
\end{equation}
larger than the bounds (\ref{tbl1}) and (\ref{tbl2}). Still, the factor (\ref
{orders}) improves the first scenario by about an order of magnitude.

\section{Composite particles}

\setcounter{equation}{0}

In the previous section we have used the dispersion relations predicted by
our models (\ref{noH}) and (\ref{mink}) for elementary particles, but we
have applied them to composite particles, such as protons and iron atoms. In
this section we investigate the dispersion relations of composite particles
and discuss some phenomenological consequences. In particular, we show that
in composite particles lower values of $b_{0}$ are favored.

A good starting point is to assume that at high energies the composite
particle can be described in a purely kinematic way, namely by constituents
moving with the same velocity $\mathbf{v}$. The effects of interactions
among constituents will not be studied in this paper. Instead, the
low-energy dispersion relation is just the relativistic one. Later we paste
it together with the high-energy dispersion relation and obtain an
approximate dispersion relation for the composite particle, valid both at
low and high energies.

We begin considering constituents with dispersion relations 
\begin{equation}
E_{i}=|\mathbf{p}_{i}|\sqrt{1+\left( \frac{\eta _{i}^{2}\mathbf{p}_{i}^{2}}{%
\Lambda _{L}^{2}}\right) ^{n-1}}.  \label{drcost}
\end{equation}
Their velocities are 
\begin{equation}
\mathbf{v}_{i}=\frac{\mathrm{d}E_{i}}{\mathrm{d}\mathbf{p}_{i}}=\frac{%
\mathbf{p}_{i}}{E_{i}}\left( 1+n\left( \frac{\eta _{i}^{2}p_{i}^{2}}{\Lambda
_{L}^{2}}\right) ^{n-1}\right) ,  \label{sq}
\end{equation}
where $p_{i}=|\mathbf{p}_{i}|$. Setting $\mathbf{v}_{i}=\mathbf{v}$ for
every $i$ it is easy to derive the dispersion relation of the composite
particle. Calling 
\begin{equation}
x_{i}=\left( \frac{\eta _{i}^{2}p_{i}^{2}}{\Lambda _{L}^{2}}\right) ^{n-1}
\label{xi}
\end{equation}
and squaring (\ref{sq}), we get the equations 
\[
v^{2}(1+x_{i})=\left( 1+nx_{i}\right) ^{2}. 
\]
Their solutions are 
\[
x_{i}=\frac{v^{2}-2n+v\sqrt{v^{2}+4n(n-1)}}{2n^{2}}\equiv x(v). 
\]
(It is easy to check that the other solution of the quadratic equation is
not acceptable). Then we have 
\[
p_{i}=\frac{x^{1/(2n-2)}}{\eta _{i}}\Lambda _{L},\qquad \mathbf{p}_{i}=%
\mathbf{v}\frac{E_{i}}{1+nx},\qquad E_{i}=\frac{x^{1/(2n-2)}}{\eta _{i}}%
\Lambda _{L}\sqrt{1+x}, 
\]
and therefore the total momentum and total energy are 
\[
\mathbf{P}=\sum_{i}\mathbf{p}_{i}=\mathbf{v}\frac{E}{1+nx},\qquad
E=\sum_{i}E_{i}=\frac{x^{1/(2n-2)}}{\eta }\Lambda _{L}\sqrt{1+x}, 
\]
where $\eta $ is defined by 
\begin{equation}
\frac{1}{\eta }=\sum_{i}\frac{1}{\eta _{i}}.  \label{resisto}
\end{equation}
Moreover, since $\mathbf{p}_{i}=\mathbf{v}p_{i}/v$, we have also 
\[
P=\sum_{i}p_{i}=\frac{x^{1/(2n-2)}}{\eta }\Lambda _{L},\qquad x=\left( \frac{%
\eta ^{2}\mathbf{P}^{2}}{\Lambda _{L}^{2}}\right) ^{n-1}. 
\]
Thus, we find that $E$ and $\mathbf{P}$ are related by the collective
dispersion relation 
\[
E=|\mathbf{P}|\sqrt{1+\left( \frac{\eta ^{2}\mathbf{P}^{2}}{\Lambda _{L}^{2}}%
\right) ^{n-1}}, 
\]
which has the same form as the dispersion relations (\ref{drcost}) of the
constituents.

The crucial result is the composition rule (\ref{resisto}), which states
that ``the weakest wins'', namely if one constituent has a $\eta _{\bar{%
\imath}}$ much smaller than the $\eta _{i}$'s of the other constituents,
then the composite particle has a $\eta $ practically equal to $\eta _{\bar{%
\imath}}$.

Note that for $n=0$, $m_{i}=\Lambda _{L}/\eta _{i}$, we get the dispersion
relation of relativistic theories, with the usual composition rule for the
mass, namely $\sum_{i}m_{i}=\sum_{i}\Lambda _{L}/\eta _{i}=\Lambda _{L}/\eta
=M$.

The result just found can be extended to more general dispersion relations
of the form 
\begin{equation}
E_{i}=|\mathbf{p}_{i}|f\left( x_{i}\right) ,\qquad x_{i}=\left( \frac{\eta
_{i}^{2}\mathbf{p}_{i}^{2}}{\Lambda _{L}^{2}}\right) ^{n-1}.
\label{dispegen}
\end{equation}
Squaring the velocities 
\begin{equation}
\mathbf{v}_{i}=\frac{\mathrm{d}E_{i}}{\mathrm{d}\mathbf{p}_{i}}=\frac{%
\mathbf{p}_{i}}{|\mathbf{p}_{i}|}\left( f+2(n-1)x_{i}f^{\prime }\right)
\label{velgen}
\end{equation}
and equating them to $\mathbf{v}$, we get the equations 
\[
v^{2}=\left( f(x_{i})+2(n-1)x_{i}f^{\prime }(x_{i})\right) ^{2}. 
\]
Assume that the solution is unique, $x_{i}=x(v)$. Then, proceeding as above,
we easily find that $E$ and $\mathbf{P}$ are related by the collective
dispersion relation 
\[
E=|\mathbf{P}|f(x),\qquad x=\left( \frac{\eta ^{2}\mathbf{P}^{2}}{\Lambda
_{L}^{2}}\right) ^{n-1}, 
\]
where $\eta $ is still given by (\ref{resisto}). Again, the dispersion
relation of the composite particle has the same form as the dispersion
relations of its constituents.

Although the procedure just outlined is general, few dispersion relations
can be treated so simply. More complicated relations generate polynomial
equations of high degree, and the dispersion relation of the composite
particle does not have the form of the dispersion relations of its
constituents. To convince oneselves of this, it is sufficient to repeat the
derivation adding mass terms to (\ref{drcost}) and (\ref{dispegen}). Yet,
masses are important for the Cherenkov effect, because they determine the
energy threshold. To apply our results to ultrahigh-energy cosmic rays we
argue as follows.

The dispersion relations (\ref{drcost}) and (\ref{dispegen}) are good
approximations at high energies, namely when the Lorentz violating
corrections start to become important and the mass becomes negligible with
respect to them. These are precisely the energies above threshold. Indeed,
the emission of radiation is the first effect of the Lorentz violation in
the phenomenon we are considering. Instead, at energies much smaller than
the threshold the Lorentz violating corrections become negligible with
respect to the mass, and the usual relativistic dispersion relation $E=\sqrt{%
M^{2}+p^{2}}$ holds, where $M$ is the mass of the composite particle. The
full dispersion relation of the composite particle can be well approximated
pasting the low- and high-energy dispersion relations.

Now, consider ultrahigh-energy cosmic rays. In our model, setting all
non-crucial parameters but the mass to zero as in (\ref{seconda}), or
relating the parameters as in (\ref{aut}), at high energies quarks have
dispersion relations (\ref{dispegen}) with $n=2$, $f(x)=1+x$, or (\ref
{drcost}) with $n=3$, respectively. In both cases $\eta _{i}^{2}=|b_{0i}|$.
Thus, the dispersion relation of the composite particle can be approximated
by the formulas 
\[
E=\sqrt{M^{2}+p^{2}\left( 1+\eta ^{2}\frac{p^{2}}{\Lambda _{L}^{2}}\right)
^{2}},\qquad E=\sqrt{M^{2}+p^{2}+\eta ^{4}\frac{(p^{2})^{3}}{\Lambda _{L}^{4}%
}}, 
\]
in the first and second scenarios, respectively, where $\eta $ is determined
by equation (\ref{resisto}).

Let us illustrate some basic properties of the composition rule (\ref
{resisto}). Consider the proton. Its dispersion relation has the same form
as the dispersion relations of its constituents, with 
\[
|b_{0p}|=\eta ^{2}=\left( \frac{2}{|b_{0u}|^{1/2}}+\frac{1}{|b_{0d}|^{1/2}}%
\right) ^{-2}, 
\]
where $b_{0u}$ and $b_{0d}$ are the $b_{0}$-parameters of the quarks $u$ and 
$d$, respectively. If $|b_{0d}|\ll |b_{0u}|$ then $|b_{0p}|\sim |b_{0d}|$,
while if $|b_{0u}|\ll |b_{0d}|$ then $|b_{0p}|\sim |b_{0u}|/4$. This means
that in composite particles, smaller values of $|b_{0}|$ are favored and the
energy threshold for Cherenkov radiation is enhanced. In practice,
compositeness creates a sort of screening for the emission of radiation and
makes it easier to justify the small numbers found in the previous section.

We have no reason to assume that $|b_{0u}|$ and $|b_{0d}|$ are of the same
orders. Let us first assume $|b_{0u}|\gg |b_{0d}|$ and normalize $\tau _{0}$
to one, as usual. Then, if ultrahigh-energy cosmic rays are protons we have 
\[
|b_{0p}|\sim |b_{0d}|, 
\]
while if they are iron atoms we gain an extra factor 7396: 
\[
b_{0\text{iron}}=\left( \frac{82}{|b_{0u}|^{1/2}}+\frac{86}{|b_{0d}|^{1/2}}%
\right) ^{-2}\sim \frac{|b_{0d}|}{7396}. 
\]
In the first scenario described in the previous section the observation of
ultrahigh-energy cosmic rays made of iron atoms can be explained with 
\[
b_{0d}=4.1\cdot 10^{-12+2k}\qquad \text{for }\Lambda _{L}=10^{14+k}\text{%
GeV,} 
\]
and $b_{0d}=1$ for $\Lambda _{L}=4.9\cdot 10^{19}$GeV.

We see that when the composite structure gets more complex it becomes easier
to generate small number from larger ones. Patterns like e.g. 
\begin{eqnarray*}
\tau _{0} &=&1,\qquad b_{0u}\sim 10^{-6},\qquad b_{0d}\sim 4\cdot
10^{-12},\qquad \Lambda _{L}\sim 10^{14}\text{GeV,} \\
\tau _{0} &=&1,\qquad b_{0u}\sim 10^{-3},\qquad b_{0d}\sim 4\cdot
10^{-6},\qquad \Lambda _{L}\sim 10^{17}\text{GeV,}
\end{eqnarray*}
are compatible with a scale of Lorentz violation smaller than the Planck
scale. The values of $b_{0u}$ have been chosen to lie somewhere in the
middle between those of $\tau _{0}$ and those of $b_{0d}$ for illustrative
purposes.

In the second scenario we can gain an extra factor (\ref{orders}) and can
explain the same $b_{0\text{iron}}$'s with slightly larger $b_{0d}$'s: 
\begin{eqnarray*}
\tau _{0} &=&1,\qquad b_{0u}\sim 10^{-6},\qquad b_{0d}\sim 5\cdot
10^{-11},\qquad \Lambda _{L}\sim 10^{14}\text{GeV,} \\
\tau _{0} &=&1,\qquad b_{0u}\sim 10^{-3},\qquad b_{0d}\sim 3\cdot
10^{-5},\qquad \Lambda _{L}\sim 10^{17}\text{GeV,}
\end{eqnarray*}

Finally, if we assume $b_{0u}\sim b_{0d}$ we gain another factor 4: 
\[
|b_{0p}|\sim |b_{0n}|\sim \frac{|b_{0u}|}{9},\qquad b_{0\text{iron}}\sim 
\frac{|b_{0u}|}{28224}. 
\]

\section{Cherenkov radiation of neutrons and neutrinos}

\setcounter{equation}{0}

We know that when Lorentz symmetry is violated, several otherwise forbidden
phenomena are allowed. In this section we describe the Cherenkov radiation
of neutral particles. Because of the Lorentz violation, photon emission is
allowed by kinematics. Moreover, our models contain Pauli-like terms at the
fundamental level, which couple neutral particles to the electromagnetic
field. We take neutrons and neutrinos and consider both the Cherenkov
radiation in a medium and the effects of higher-derivative terms.

The neutron Lagrangian is 
\begin{eqnarray*}
\mathcal{L}_{\text{neutron}} &=&\mathcal{L}_{F}+\bar{\psi}_{n}\left( i\gamma
^{0}\partial _{0}+\frac{ib_{0n}}{\Lambda _{L}^{2}}\bar{\partial}{\!\!\!\!\hspace{1.1truept}\slash%
}\,^{3}+ib_{1n}\bar{\partial}\!\!\!\!\hspace{1.1truept}\slash -m_{n}-\frac{b_{n}^{\prime }}{%
\Lambda _{L}}\bar{\partial}{\!\!\!\!\hspace{1.1truept}\slash}\,^{2}\right) \psi _{n} \\
&&+\frac{e}{\Lambda _{L}}\bar{\psi}_{n}\hspace{0.02in}\left( b_{n}^{\prime
\prime }\sigma _{ij}F^{ij}+\frac{b_{0n}^{\prime }}{\Lambda _{L}}\gamma
_{i}\partial _{j}F_{ij}\right) \psi _{n}+ie\frac{b_{0n}^{\prime \prime }}{%
2\Lambda _{L}^{2}}F_{ij}\left( \bar{\psi}_{n}\gamma _{i}\hspace{0.02in}%
\overleftrightarrow{{\bar{D}}}_{j}\psi _{n}\right)
\end{eqnarray*}
and the kinematics of the Cherenkov process is the one of section 4. The
Cherenkov radiation can be studied adapting the results found for the
proton. Indeed, after replacements of the form 
\begin{equation}
b^{\prime \prime }=\frac{\tilde{b}^{\prime \prime }}{e},\qquad b_{0}^{\prime
}=\frac{\tilde{b}_{0}^{\prime }}{e},\qquad b_{0}^{\prime \prime }=\frac{%
\tilde{b}_{0}^{\prime \prime }}{e},\qquad  \label{repl}
\end{equation}
the neutron Lagrangian matches the proton Lagrangian at $e=0$. As far as the
Cherenkov radiation in a medium is concerned, we must evaluate formulas up
to $\mathcal{O}(1/\Lambda _{L}^{2})$ corrections, perform the replacements (%
\ref{repl}), followed by the limit $e\rightarrow 0$ and the converse
replacements. We find 
\[
\frac{\mathrm{d}v}{\mathrm{d}t}=-\frac{16\alpha m_{n}^{3}\mu
^{5}b_{n}^{\prime \prime \hspace{0.01in}2}(nv-1)^{4}}{15\Lambda
_{L}^{2}n^{8}(n^{2}-1)^{5}v^{2}\sqrt{1-v^{2}}}\left[
(5n^{2}-1)(6n^{2}+1)-4n(n^{2}+1)v-5n^{2}(5n^{2}-1)v^{2}\right] , 
\]
with the velocity $v$ defined as in (\ref{vE}).

We can also make an analysis similar to the one of section 5. In the limit $%
b_{0n}^{2}\ll \tau _{0}$, the analogue of (\ref{radti0}) gives the typical
radiation time 
\[
\left. t_{fn}\right| _{b_{0n}^{2}\ll \tau _{0}}\sim \frac{\tau _{0}}{6\sqrt{5%
}b_{n}^{\prime \prime }{}^{2}|b_{0n}|\alpha m_{n}}. 
\]

In both cases we see that the Cherenkov radiation of neutrons crucially
depends on the parameter $b_{n}^{\prime \prime }$, besides $\tau _{0}$ and $%
b_{0n}$. Thus, measurements cannot say much about the scales of Lorentz
violation, which are our main interest in this paper, but can put bounds on
the values of the parameters $b_{n}^{\prime \prime }$. Nevertheless, some
aspects of the neutron Cherenkov radiation may deserve further study, since
it is known that in some cases the Lorentz violation makes protons decay
into neutrons \cite{glashow}. Then ultrahigh-energy cosmic rays could be
regarded as a mixture of protons and neutrons and both particles would
contribute to the emission of Cherenkov radiation.

If neutrinos are taken to be massive, their case is entirely analogous to
the case of the neutron. Instead, if we neglect their mass (or assume that
they are massless and that neutrino oscillations have a different
explanation) the Lagrangian that describes interactions with the
electromagnetic field is 
\[
\mathcal{L}_{\text{neutrino}}=\mathcal{L}_{F}+\bar{\nu}\left( i\gamma
^{0}\partial _{0}+\frac{ib_{0\nu }}{\Lambda _{L}^{2}}\bar{\partial}{\!\!\!%
\!\hspace{1.1truept}\slash}\,^{3}+ib_{1\nu }\bar{\partial}\!\!\!\!\hspace{1.1truept}\slash \right) \nu +\frac{%
eb_{0\nu }^{\prime }}{\Lambda _{L}^{2}}\partial _{j}F_{ij}(\bar{\nu}\hspace{%
0.02in}\gamma _{i}\nu )+ie\frac{b_{0\nu }^{\prime \prime }}{2\Lambda _{L}^{2}%
}F_{ij}\left( \bar{\nu}\gamma _{i}\hspace{0.02in}\overleftrightarrow{{%
\partial }}_{j}\nu \right) . 
\]
Here the effect is of higher order, because $b^{\prime \prime }$ is absent.
Moreover, at vanishing mass the kinematics also changes. For example, in the
case $b_{1\nu }=\tau _{2}=1$, $\tau _{1}=0$, we find the $k$-range $%
0\leqslant k\leqslant p$ and 
\[
\frac{\mathrm{d}E}{\mathrm{d}t}=-\frac{517b_{0\nu }\alpha (4b_{0\nu
}^{\prime \hspace{0.01in}2}+b_{0\nu}^{\prime \prime \hspace{0.01in}2})p^{8}}{%
10080\Lambda _{L}^{6}}. 
\]

\section{Conclusions}

\setcounter{equation}{0}

If Lorentz symmetry is violated at high energies, a variety of phenomena
that are normally forbidden can take place. The investigation of these
phenomena can help us better address the search for signs of Lorentz
violation and put bounds on the parameters of the violation. In this paper
we have focused on the Cherenkov radiation in vacuo, and explored scenarios
compatible with a scale of Lorentz violation $\Lambda _{L}$ smaller than the
Planck scale. We have worked in the realm of the minimal Standard-Model
extension that violates Lorentz symmetry at high energies, preserves CPT and
rotational invariance, contains operators of higher dimensions (in
particular, four-fermion vertices) and is renormalizable by weighted power
counting. This SM\ extension offers a framework where the set of new
parameters is large enough to describe the phenomena allowed by Lorentz
violation, but sufficiently restricted to ensure a certain degree of
predictivity.

We have studied kinematic constraints for a very general class of dispersion
relations, and found an energy threshold below which particles do not
radiate. We have computed the energy loss of particles above threshold and
verified that it is so rapid that in all cases of interest the process is
practically governed by pure kinematics. For different values of the scale $%
\Lambda _{L}$ we have studied how small certain parameters must be to have
compatibility with known experimental bounds on the Lorentz violation.
Interestingly, in composite particles ``small numbers'' are easier to
explain, because the threshold for Cherenkov radiation is enhanced by a sort
of kinematic screening mechanism.

Our analysis shows that there is still the possibility that the scale of
Lorentz violation, with preserved CPT, is smaller than Planck scale. If
confirmed, this prediction would force us to think about quantum gravity
anew.

\vskip 20truept \noindent {\Large \textbf{Acknowledgments}}

\vskip 10truept

D.Anselmi wishes to thank Xinmin Zhang and the Institute of High Energy
Physics of the Chinese Academy of Sciences, Beijing, for hospitality.
D.Anselmi is supported by the Chinese Academy of Sciences visiting
professorship for senior international scientists, grant No. 2010T2J01.

\vskip 20truept \noindent {\Large \textbf{Appendix: the simplest LVSM's}}

\vskip 10truept

\renewcommand{\theequation}{A.\arabic{equation}} \setcounter{equation}{0}

For reference, in this appendix we briefly recall the simplest LVSM's and
some of their features. The minimal scalarless CPT\ invariant LVSM
schematically reads \cite{noh} 
\begin{equation}
\mathcal{L}_{\mathrm{noH}}=\mathcal{L}_{F}+\mathcal{L}_{\mathrm{{kin}f}%
}-\sum_{I=1}^{5}\frac{1}{\Lambda _{L}^{2}}g\bar{D}\bar{F}\,(\bar{\chi}_{I}%
\bar{\gamma}\chi _{I})+\frac{Y_{f}}{\Lambda _{L}^{2}}\bar{\chi}\chi \bar{\chi%
}\chi -\frac{g}{\Lambda _{L}^{2}}\bar{F}^{3},  \label{noH}
\end{equation}
where 
\begin{eqnarray*}
\mathcal{L}_{F} &=&\frac{1}{4}\sum_{G}\left( 2F_{0i}^{G}F_{0i}^{G\hspace{%
0.01in}}-F_{ij}^{G}\tau ^{G}(\bar{\Upsilon})F_{ij}^{G\hspace{0.01in}}\right)
, \\
\mathcal{L}_{\mathrm{{kin}f}} &=&\sum_{a,b=1}^{3}\sum_{I=1}^{5}\bar{\chi}%
_{I}^{a}\hspace{0.02in}i\left( \delta ^{ab}\gamma ^{0}D_{0}-\frac{b_{0}^{Iab}%
}{\Lambda _{L}^{2}}{\bar{D}\!\!\!\!\slash}\,^{3}+b_{1}^{Iab}\bar{D}\!\!\!\!%
\slash \right) \chi _{I}^{b},
\end{eqnarray*}
are the kinetic terms of gauge fields and fermions, respectively. Bars are
used to denote space components. The ``magnetic'' components $F_{ij}$ of the
field strengths are also denoted with $\bar{F}$. Moreover, $\chi
_{1}^{a}=L^{a}=(\nu _{L}^{a},\ell _{L}^{a})$, $\chi
_{2}^{a}=Q_{L}^{a}=(u_{L}^{a},d_{L}^{a})$, $\chi _{3}^{a}=\ell _{R}^{a}$, $%
\chi _{4}^{a}=u_{R}^{a}$ and $\chi _{5}^{a}=d_{R}^{a}$, $\nu ^{a}=(\nu
_{e},\nu _{\mu },\nu _{\tau })$, $\ell ^{a}=(e,\mu ,\tau )$, $u^{a}=(u,c,t)$
and $d^{a}=(d,s,b)$. The sum $\sum_{G}$ is over the gauge groups $SU(3)_{c}$%
, $SU(2)_{L}$ and $U(1)_{Y}$, and the last three terms of (\ref{noH}) are
symbolic. Finally, $\bar{\Upsilon}\equiv -\bar{D}^{2}/\Lambda _{L}^{2}$,
where $\Lambda _{L}$ is the scale of Lorentz violation, and $\tau ^{G}$ are
polynomials of degree 2. Gauge anomalies cancel out exactly as in the
Standard Model \cite{lvsm}.

The model (\ref{noH}) does not contain elementary scalar fields, but
four-fermion vertices trigger a Nambu--Jona-Lasinio mechanism that gives
masses to fermions and gauge fields, and generate Higgs bosons at low
energies as composite fields \cite{noh,noh2}.

LVSM\ versions containing elementary scalar fields and incorporating the
usual Higgs phenomenon exist as well. They can contain the
two-scalar--two-fermion vertex $(LH)^{2}/\Lambda _{L}$ and four-fermion
vertices $(\bar{\psi}\psi )^{2}/\Lambda _{L}^{2}$ at the fundamental level.

After symmetry breaking, the vertex $(LH)^{2}/\Lambda _{L}$ gives (Majorana)
masses to the left-handed neutrinos. Since this vertex is the only
dimension-5 vertex present in the LVSM, it can be used to normalize the
scale $\Lambda _{L}$. Assuming that the dimensionless couplings in front of
it are of order one we find $\Lambda _{L}\sim $ 10$^{14}$-10$^{15}$GeV \cite
{astrumia}. Four-fermion vertices can describe proton decay. The existing
bounds on proton decay can also be used to constrain $\Lambda _{L}$, and
give $\Lambda _{L}\geqslant $10$^{15}$GeV.

Other prescriptions to normalize $\Lambda _{L}$ are considered in the paper.


\begin{thebibliography}{99}
\bibitem{kostelecky}  V.A. Kosteleck\'{y} and N. Russell, \textit{Data
tables for Lorentz and CTP violation}, arXiv:0801.0287 [hep-ph].

\bibitem{halat}  D. Anselmi and M. Halat, Renormalization of Lorentz
violating theories, Phys. Rev. D 76 (2007) 125011 and arXiv:0707.2480
[hep-th].

\bibitem{LVgauge1}  D. Anselmi, Weighted power counting and Lorentz
violating gauge theories. I: General properties, Ann. Phys. 324 (2009) 874
and arXiv:0808.3470 [hep-th].

\bibitem{LVgauge2}  D. Anselmi, Weighted power counting and Lorentz
violating gauge theories. II: Classification, Ann. Phys. 324 (2009) 1058 and
arXiv:0808.3474 [hep-th].

\bibitem{lvsm}  D. Anselmi, Weighted power counting, neutrino masses and
Lorentz violating extensions of the Standard Model, Phys. Rev. D 79 (2009)
025017 and arXiv:0808.3475 [hep-ph].

\bibitem{noh}  D. Anselmi, Standard Model Without Elementary Scalars And
High Energy Lorentz Violation, Eur. Phys. J. C 65 (2010) 523 and
arXiv:0904.1849 [hep-ph].

\bibitem{gagnon}  O. Gagnon and G. Moore, Limits on Lorentz violation from
highest energy cosmic rays, Phys. Rev. D 70 (2004) 065002 and
arXiv:hep-ph/0404196.

\bibitem{taiuti}  D. Anselmi and M. Taiuti, Renormalization of high-energy
Lorentz violating QED, Phys. Rev. D 81 (2010) 085042 and arXiv:0912.0113
[hep-ph].

\bibitem{altschul}  B.D. Altschul, Finite duration and energy effects in
Lorentz-violating vacuum Cerenkov radiation, Nucl. Phys. B796 (2008) 262 and
arXiv:0709.4478 [hep-th].

\bibitem{glashow}  S.R. Coleman and G.L. Glashow, High-energy tests of
Lorentz invariance, Phys. Rev. D59 (1999) 116008 and arXiv:hep-ph/9812418.

\bibitem{vari}  R. Lehnert and R. Potting, Vacuum Cerenkov radiation, Phys.
Rev. Lett. 93 (2004) 110402 and arXiv:hep-ph/0406128;

R. Lehnert and R. Potting, The Cerenkov effect in Lorentz-violating vacua,
Phys. Rev. D70 (2004) 125010, Erratum-ibid. D70 (2004) 129906 and
arXiv:hep-ph/0408285;

B.D. Altschul, Vacuum Cerenkov radiation in Lorentz-violating theories
Without CPT Violation, Phys. Rev. Lett. 98 (2007) 041603 and
arXiv:hep-th/0609030;

B.D. Altschul, Cerenkov radiation in a Lorentz-Violating and birefringent
vacuum, Phys. Rev. D75 (2007) 105003 and arXiv:hep-th/0701270.

\bibitem{klinkhamer}  F.R. Klinkhamer and M. Schreck, New two-sided bound on
the isotropic Lorentz-violating parameter of modified Maxwell theory, Phys.
Rev. D 78 (2008) 085026 and arXiv:0809.3217 [hep-ph].

\bibitem{jackson}  J.D. Jackson, \textit{Classical electrodynamics}, \S %
14.9, John Wiley \& Sons Inc., New York, 1962.

\bibitem{francesi}  D.J. Bird et al., Detection of a cosmic ray with
measured energy well beyond the expected spectral cutoff due to cosmic
microwave radiation, Astrophys. J. 441 (1995) 144.

\bibitem{jones}  D.R.T. Jones, Two-loop $\beta $ function for a $G_{1}\times
G_{2}$ gauge group, Phys. Rev. D 25 (1982) 581.

\bibitem{mz}  F. Jegerlehner, The running fine structure constant $\alpha (E)
$ via the Adler function, Nuc. Phys. B - Proc. Supp. 181 (2008) 135 and
arXiv:0807.4206 [hep-ph].

\bibitem{pdg}  K. Nakamura et al. (Particle Data Group), J. Phys. G 37
(2010) 075021.

\bibitem{noh2}  D.Anselmi and E. Ciuffoli, Low-energy phenomenology of
scalarless Standard-Model extensions with high-energy Lorentz violation,
Phys. Rev. D 83 (2011) 056005 and arXiv:1101.2014 [hep-ph].

\bibitem{astrumia}  See for example, A. Strumia and F. Vissani, Neutrino
masses and mixings and..., arXiv:hep-ph/0606054 [hep-ph].
\end{thebibliography}
\end{document}